# Window Filtering Algorithm for Pulsed Light Coherent Combining of Low Repetition Frequency

Jiali Zhang, Jie Cao, Qun Hao, Yang Cheng, Liquan Dong, Bin Han, and Xuesheng Liu

*Abstract*—The multi-dithering method has been well verified in phase locking of polarization coherent combination experiment. However, it is hard to apply to low repetition frequency pulsed lasers, since there exists an overlap frequency domain between pulse laser and the amplitude phase noise and traditional filters cannot effectively separate phase noise. Aiming to solve the problem in this paper, we propose a novel method of pulse noise detection, identification, and filtering based on the autocorrelation characteristics between noise signals. In the proposed algorithm, a self-designed window algorithm is used to identify the pulse, and then the pulse signal group in the window is replaced by interpolation, which effectively filter the pulse signal doped in the phase noise within 0.1 ms. After filtering the pulses in the phase noise, the phase difference of two pulsed beams (10 kHz) is successfully compensated to zero in 1 ms, and the coherent combination of closed-loop phase lock is realized. At the same time, the phase correction times are few, the phase lock effect is stable, and the final light intensity increases to the ideal value (0.9 Imax).

## I. Introduction

Low repetition frequency (LRF) nanosecond pulsed laser light sources have a wide range of applications such as precision machining, photoelectric detection, and lidar [1-3]. Compared with other lasers, fiber lasers [4-7] have many unique advantages, such as narrow linewidth, high peak power, good beam quality, and easy beam combination, which promote the development of fiber pulsed optical coherent combination [8-10] technology based on the main oscillating power amplifier (MOPA) structure [11,12]. The key technology to realize the coherent combination of pulsed beams is the precise control of phase, which ensures the in-phase output of each pulsed laser. Active phase-locking schemes based on MOPA structure mainly include the heterodyne method [13,14], the stochastic parallel gradient descent (SPGD) method [15,16], dithering method [17-20], etc. Because coherent beam combination (CBC) based on multi-dithering method only needs one photodetector, the system structure is simple, and the control bandwidth is mainly determined by the speed of the processing circuit, it is considered to be a promising method for obtaining high brightness laser [21-24].

At present, in the phase control of most CBC, the phase fluctuation of each laser is generally calculated by the power fluctuation of the combined beam, and then the phase difference between the lasers is corrected to achieve the in-phase output of the laser [9,16]. However, the laser pulse itself is a kind of intensity fluctuation, which will inevitably affect the phase control of CBC. Especially for low-repetition pulse lasers, because the frequency domain of the laser overlaps the frequency domain of the phase noise [25], the traditional filtering method cannot effectively filter the laser pulse and retain the phase noise.

At present, the multi-dithering algorithm can only compensate phase of the high-frequency pulsed beam. Daniault et al actively coherent combined two femtosecond pulsed lasers with repetition rate of more than 100 kHz [26]. Active coherent combining of two nanosecond laser pulses with repetition rate of 25 kHz was reported by Daniault et al [10]. However, for the coherent combination of LRF (≤10kHz) pulsed laser beams, when the customized band-stop filter with fixed cut-off frequency removes the pulse noise, the details of the signal is blurred, which leads to the inability to extract the signal containing phase noise and achieve the coherent combination of LRF pulsed laser beams [9,10]. The reason is that the pulse repetition frequency (PRF) is at the same order of magnitude as the phase noise, and the signal received by the photodetector is the small-value phase noise signal plus large-value pulse signal, and the phase noise signal with low amplitude will be greatly reduced during filtering. Therefore, an adaptive pulse filtering method is urgently needed to remove the LRF pulse noise as well as protect the details of the phase noise signal, to further complete the coherent combination of LRF pulse beams.

In this paper, the LRF pulse noise detection and filtering method proposed can automatically adapt to the variation of amplitude and broadening of pulse noise. It deals with the phase

This work was supported in part by the funding of foundation enhancement program under Grant (2019-JCJQ-JJ-273); National Natural Science Foundation of China (61871031, 61875012, 61905014).

Jiali Zhang, Jie Cao, Qun Hao, Yang Cheng, Liquan Dong, and Bin Han are with the School of Optics and photonics, Beijing Institute of Technology, Beijing, 100081, China.

Xuesheng Liu is with the Beijing Engineering Research Center of Laser Technology, Beijing University of Technology, Beijing, 100124, China.
*Corresponding author: qhao@bit.edu.cn and liuxuesheng@bjut.edu.cn

noise pollution points but does not change the non-pollution signal points. Not only has a good LRF pulse noise removal ability, but also good protection of phase noise signal details. In the experiment of CBC, this method eliminates the band-stop filter directly, which not only reduces the complexity of the whole experimental system but also saves a lot of time spent by researchers in adjusting the experimental system.

## II. THEORETICAL ANALYSIS

Multi-dithering coherent combining algorithm is to mark phase noise by applying a modulation signal to the beam. Then the phase error signal of each beam is separated from the detected photoelectric signal, which is fed back to the phase modulator as the control signal to achieve the laser phase locking of each beam. The output current of the photodetector in the coherent combining system can be expressed as [27]

$$i_{PD}(t) = R_{PD} \cdot S \cdot P(t), \quad (1)$$

where $R_{PD}$ represents the responsivity of the photodetector, $S$ represents the photodetector area and $P(t)$ represents the optical power of the combined beam on the photodetector.

The power fluctuation of the combined beam represents the phase fluctuation of each laser, and the phase noise is extracted by coherent detection to realize the phase compensation. However, the laser pulse itself is a kind of light intensity fluctuation, and the multi-dithering method can only effectively perform phase demodulation after filtering it.

Suppose the pulsed laser light field involved in the combining be expressed as [28]

$$E_m(x, y, z, t) = A_m(x, y, z)\Gamma_m(t)e^{-j[\omega t - \varphi_m(t)]}, \quad (2)$$

where $A_m(x, y, z)$ is the spatial term of the beam amplitude, and $\Gamma_m(t)$ is the amplitude time term of the $m$th beam.

Further assuming that the pulse waveforms of each channel are consistent and synchronized in the time domain, and only considering the influence of pulse periodic amplitude changes on phase control, then the light intensity distribution of the combined beam is [28]

$$I(x,y,z,t) = I_0(t)(\sum_{m=1}^{M} A_{m_i}(x,y,z,t))(\sum_{m=1}^{M} A_{n_i}(x,y,z,t))^*$$
$$= I_0(t)[\sum_{m=1}^{M} I_m(x,y,z) + \sum_{m_i \neq m_j} A_{0m_i}(x,y,z)A^*_{0m_j}(x,y,z)e^{j(\varphi_{m_i}(t) - \varphi_{m_j}(t))}], \quad (3)$$

where $I_m(x, y, z) = |A_m(x, y, z)|^2$ is the spatial distribution of coherent combined light intensity.

Since $I_0(t)$ is a function of the pulse repetition frequency ($f_{RR}$) as the period, $I_0(t)$ can be transformed into an even function by shifting in the time domain, and Fourier expansion of it will give [28]

$$I_0(t) = \frac{a_0}{2} + \sum_{k=1}^{\infty} a_k \cos(2\pi k f_{RR} t). \quad (4)$$

Taking into account the Fourier transform of the cosine function is [28]

$$\mathbb{F}[\cos(2\pi k f_0 t)] = \frac{1}{2}[\delta(f - f_0) + \delta(f + f_0)], \quad (5)$$

then the spectrum of the time domain $I_0(t)$ of the coherent combined light intensity is [28]

$$F_1 = \mathbb{F}[I_0(t)] = \frac{a_0}{2}\delta(0) + \frac{1}{2}\sum_{k=1}^{\infty}[\delta(f - kf_{RR}) + \delta(f + kf_{RR})]. \quad (6)$$

It can be seen from formula (6) that the spectrum of $I_0(t)$ is a separated spectrum with an interval of $f_{RR}$. In Eq. (3), the phase term in the square brackets not only contains the external phase noise, but also contains the small disturbance imposed by the control algorithm on the modulator. According to the convolution theorem, the spectrum ($F$) of $I(x, y, z, t)$ is the convolution of the spectrum ($F_1$) of $I_0(t)$ with the spectrum ($F_2$) of the phase term. Since the convolution of the $\delta$ function and the ordinary function satisfies: $f(x) * \delta(x - x_0) = f(x - x_0)$, The spectrum $F$ of $I(x, y, z, t)$ can be expressed as [28]

$$F = \frac{a_0}{2}F_2 + \frac{1}{2}\sum_{k=1}^{\infty} a_k[F_2(f - kf_{RR}) + F_2(f + kf_{RR})]. \quad (7)$$

Remove the negative frequency components that have no physical meaning. When the cutoff frequency of $F$ ($f_M$) is less than $f_{RR}$, the information of $F_2$ can be completely preserved in the frequency domain. However, when the cutoff frequency ($f_M$) of $F$ is greater than $f_{RR}$, the information of $F_2$ cannot be completely preserved in the frequency domain.

According to the relationship between the cutoff frequency ($f_D$) of algorithm disturbance and repetition frequency ($f_{RR}$) of the pulsed laser, it can be divided into two cases: (1) When $f_D < f_{RR}$, a low-pass filter with cutoff frequency between $f_M$ and $f_{RR}$ can be used to filter out the frequency components higher than $f_{RR}$ in the light intensity variation. The light intensity variation after low-pass filtering do not contain the light intensity fluctuation introduced by the pulse laser, and only the light intensity variation caused by phase noise and algorithm disturbance are retained. (2) When $f_D \geq f_{RR}$, the light intensity change caused by phase noise cannot be extracted by low-pass filter [28].

In the second case, when the laser pulse repetition rate is equivalent to the phase noise frequency. In this paper, the window filter algorithm is used to filter and interpolate the pulsed laser, which effectively protects the original phase noise and achieve the active phase control. In the coherent combination system, in the absence of pulse noise interference, the phase noise signal locally appears to have similar signal values. The smaller the distance of the sampling point, the higher the correlation degree of the signal, and the smaller the difference of phase noise. At this time, the signal value difference between two adjacent phase noise must be less than a certain value. When the pulse appears, its amplitude will be far away from the signal value in the neighborhood, and the pulse occupies a relatively small isolated area, which is used to judge the pulse point from the signal amplitude. Assuming that $y_{i-1}$, $y_i$, $y_{i+1}$ are the signal values corresponding to three adjacent signal sampling points of $i-1$, $i$, $i+1$, respectively, the relationship between the amplitudes of the three adjacent signals is defined as:

$$k = \left| \frac{y_i - y_{i+1}}{y_i - y_{i-1}} \right| \leq Th \ (i = 1, 2...n-1) \qquad (8)$$

Equation (8) represents the autocorrelation of the signal, where $k$ represents the changing rate of the noise amplitude, which can be determined by the actual conditions of different experimental environments, and $Th$ represents the threshold. If formula (8) is true, it means that the signal point $y_i$ is not polluted by the pulse noise, otherwise, it means that the signal point $y_i$ is polluted by the pulse.

The above formula is derived on the basis that the amplitude and waveform of the pulsed light are exactly the same, but in fact the pulsed laser can be distorted during amplification and transmission and affected by the uncertain experimental environment, there will be a sudden change in the amplitude. When the amplitude is close to the amplitude of the phase noise, it is difficult to determine whether the mutation point is a phase noise signal or a pulse signal, so it is necessary to further distinguish the pulse noise point and the phase noise point from the width. After the LRF pulsed light is sampled by the high-frequency AD module, the pulse point appears as a large amplitude isolated point. Consider the pulse light repetition frequency $f_{rep}$, the pulse width $\tau$, and the sampling rate $f_s$ AD module as an example. Assuming that the pulse is expanded by $\tau_1$ due to amplification and other factors, then a pulse can be sampled to the $f_s \tau_1$ point theoretically. In the experimental environment, considering the existence of missing signal points. Therefore, we stipulate that the number of points collected is $\beta \cdot f_s \tau_1$, and the value range of the coefficient $\beta$ here is recorded as (0.8~1). If the amplitudes of the $\beta \cdot f_s \tau_1$ noise points that are continuously collected are all within the voltage threshold range set in advance, it indicates that the collected signal is an impulse noise point. If only the amplitudes of individual points of the $\beta \cdot f_s \tau_1$ noise points collected continuously are within the preset voltage threshold range, it means that the collected signal points are phase noise points. If the width of the sampling point is further determined to be a pulse noise, we will set a filter window to achieve filtering. The window filter algorithm first sets the size of the filter window $d$ according to the pulse width $\tau$, the filter window width needs to be greater than the pulse width, then detects the pulse, and set it as the initial point, and presets the window where the next pulse appears by the pulse period. When the window arrives, the interpolated signal is used to replace the pulse signal in the window to filter out the pulse noise, so that the impact of the pulse noise signal on demodulation of the phase noise signal is minimized. The proportion of the replaced signal in the entire integration period is so small that it can be ignored. At the same time, in order to prevent the pulse from drifting outside the window, when the detected pulse is not within the window, the window needs to be adjusted in time before filtering. Therefore, this method can achieve LRF pulse light filtering without affecting the final phase lock. To reflect that the algorithm has an adaptability to change in noise amplitude and width more intuitively, we use the algorithm flow chart to express it as given below:

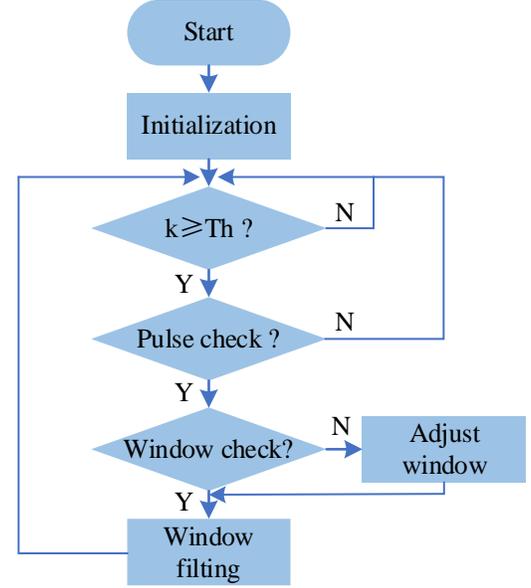

Fig. 1. Window filtering flowchart

### III. SIMULATION AND RESULT ANALYSIS

To verify that the multi-dithering method based on window filtering can realize the coherent combination of LRF pulse beams, numerical simulation was carried out for this purpose. The number of beams involved in the combination is 2 channels, the PRF is set to 10 kHz, which is at the same frequency order as the large amplitude-phase noise of the experimental environment. The frequency of phase noise is mainly concentrated within 5 kHz [25,29], and its maximum amplitude is $\lambda/20$ [29]. The pulse width is 10 ns, and the sampling frequency of the AD module is 10 MHz, and the large phase noise frequency distribution is less than 10 kHz.

#### A. Beam combination effect when there is pulse noise

In order to build a more accurate phase noise, based on the Fourier transform principle, a model of sine wave superposition and mixed random white noise is used to build the phase noise model without pulse signal pollution, as shown in equation (8), and its time domain and frequency-domain are shown in Fig. 2.

$$N_{noise} = \sum_{i=1}^{n} A_i \sin(\omega_i t + \varphi_i) + n_{noise} \qquad (9)$$

Where, $A_i$, $\omega_i$ and $\varphi_i$ are the amplitude, frequency and initial phase of the sine wave, respectively. In order to be consistent with the actual environmental noise, these parameters are generated by random functions. $n_{noise}$ is random white noise, which is directly invoked by software and is mainly used to simulate high-frequency noise.

observation, we amplify the phase noise with a frequency lower than 500 kHz, as shown in Fig. 2(c). However, for fiber laser amplification systems, the source of phase noise is mainly caused by mechanical vibration, and the amplitude of phase noise is basically maintained above -80dB (frequency below 10kHz), as shown in Fig. 2(d).

Then, the phase noise is respectively superimposed on the two pulsed optical signals to simulate the beam combination of the two pulses in the open loop. The intensity distribution of the combined pulse light in the time domain and the frequency-domain is shown in Fig. 3(a) and 3(b).

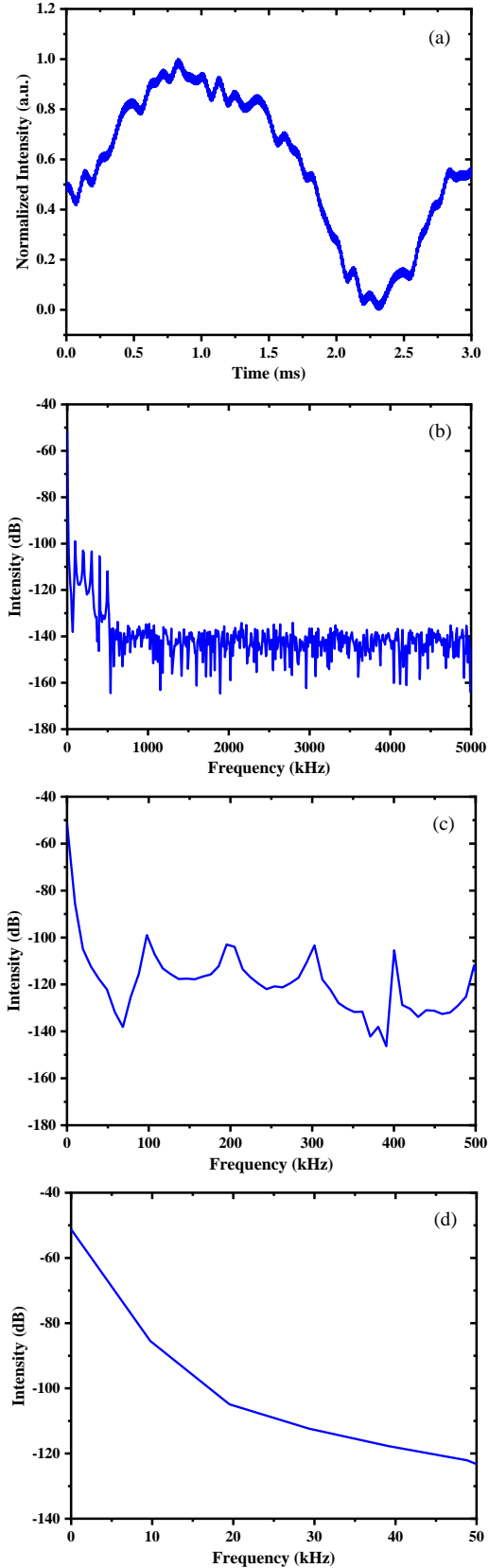

Fig. 2. The intensity of noise in time-domain and frequency-domain. (a) Time-domain phase noise; (b), (c), (d) Frequency-domain phase noise.

It can be seen from Fig. 2(b) that the frequency corresponding to large amplitude-phase noise intensity (above -140 dB) is lower than 500 kHz. In order to facilitate

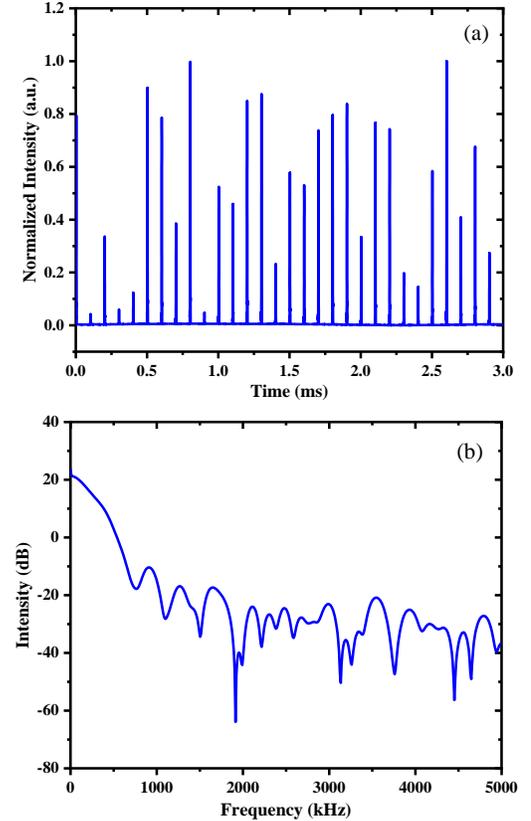

Fig. 3. The intensity of combined pulse light in the time domain and frequency-domain when open loop. (a) Time-domain; (b) Frequency-domain.

From the timing distribution of the combined pulse light as shown in Fig. 3(a), it can be seen that the intensity changes up and down in the open loop. From the frequency-domain characteristics of the combined pulse light intensity in Fig. 3(b), it can be seen that the maximum value of the pulsed intensity after combination in the open loop has reached more than 20dB. The reason lies in the introduction of pulsed light signals, resulting in a strong light intensity value in the frequency spectrum near 10 kHz.

### B. Beam combination effect after pulse noise is filtered out

The multi-dithering method based on window filtering first needs to identify the pulse signal then set the window, and finally use the interpolation method to replace the pulse signal group in the window, to filter out the pulse and effectively protect the phase noise. The LRF pulsed laser signal has the characteristics of sudden change, large amplitude, periodicity, isolated point, and so on. Using these characteristics, the algorithm first continuously collects n-period pulsed light signals and marks the signal points with sudden changes and large amplitudes as hypothetical pulse points. According to the periodicity of the pulsed light, a filter window is set in the algorithm, and the window width (100ns) is 10 times the pulse width (10ns), as shown in Fig. 4. Specific steps are: take the currently recognized hypothetical pulse point as the starting point, take the repetition frequency of the pulse as the period, and set 5 windows in turn. When a sudden change is detected in the 5 windows, the assumed pulse point is established, and the signal in the window is replaced by an interpolation signal at this time to complete the filtering, as shown in Fig. 5(a).

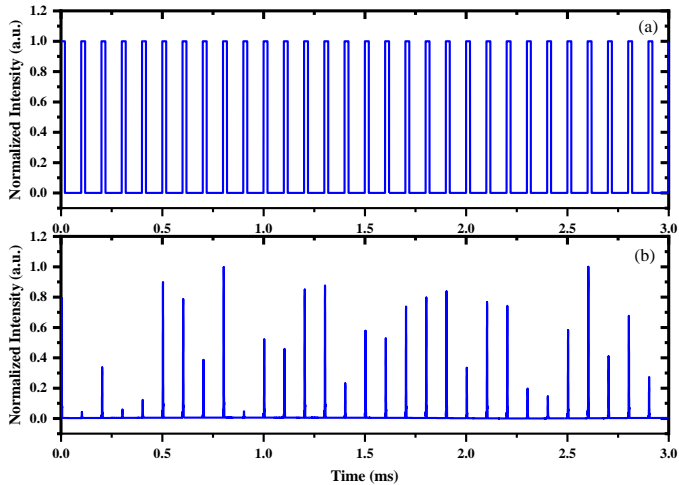

Fig. 4. The filter window and combined pulse light in the open loop. (a) Filter window; (b) Combined pulse light in the open loop.

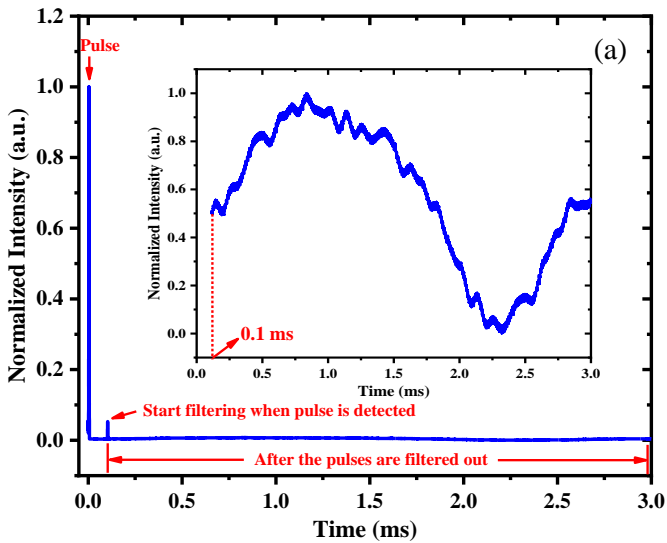

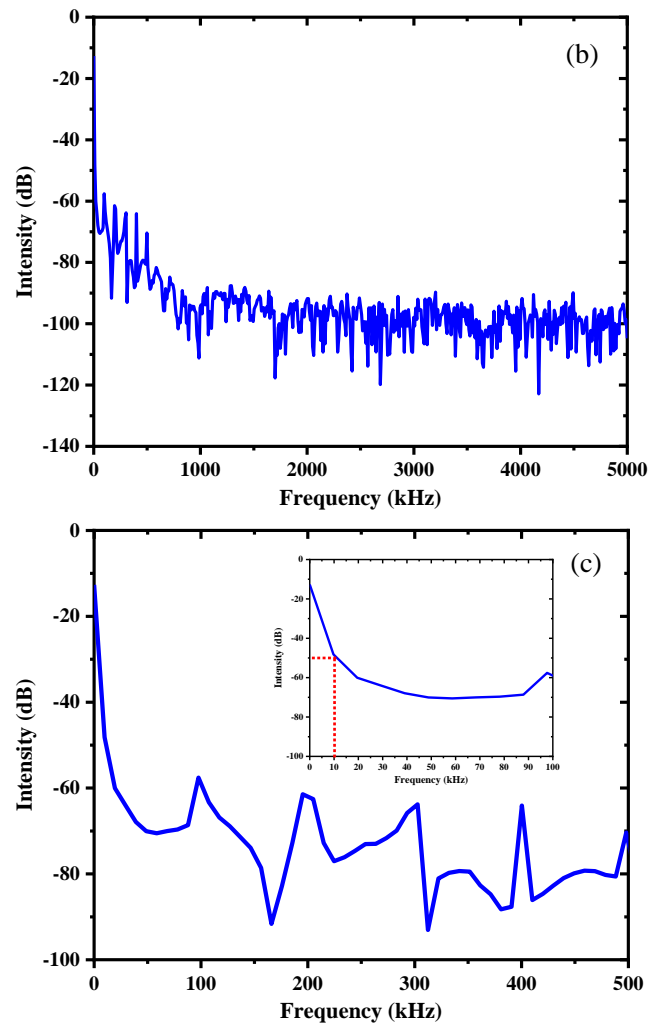

Fig. 5. The intensity changes of phase noise in time domain and frequency-domain after the pulses are filtered out in the open loop. (a) Time-domain phase noise; (b), (c) Frequency-domain phase noise.

It can be seen from Fig. 5(a) that after the window filter algorithm is turned on, the filter window detects pulses within 0.1 ms and starts to filter them. After the pulses are filtered out and replaced by interpolation, only the phase noise is retained, as shown in the inset of Fig. 5(a). Fourier transform was performed on the phase noise after filtering the pulse, as shown in Fig. 5(b). In order to more intuitively observe the change of large-value phase noise in the frequency domain, the phase noise with a frequency lower than 500 kHz in Fig. 5(b) was broadened, as shown in Fig. 5(c). Fig. 5(c) is compared with Fig. 3(b), the maximum value of the noise intensity in the frequency domain has dropped from 23 dB to -12 dB. Especially when the noise frequency is 10 kHz, the noise intensity decreases from 20 dB to -50 dB. The above results indicate that the large pulsed light frequency components in the frequency-domain have been removed.

After the pulse is filtered out, the phase noise signal will be fed the multi-dithering algorithm. During the implementation of the multi-dithering algorithm, window filter still needs to work all the time and adjust the window. When the window fails to detect a sudden high signal multiple times, it is necessary to re-identify the pulse and adjust the window. Stop adjusting the window until the light pulse signal is detected. After the pulse signal is filtered, the phase-locked coherent combination of two pulse beams can be realized, and the phase-locked effect is shown in Fig. 6. The inset in Fig. 6 shows that the phase difference of the two pulse beams in the phase-locking process. Fig. 7 is the time-domain diagram of the coherent combination of the two pulsed beams (PRF of 10 kHz) during the phase-locking process.

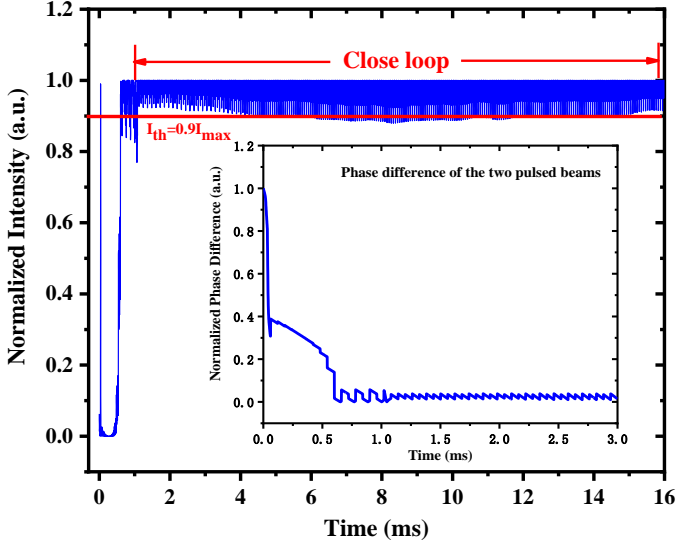

Fig. 6. The intensity of the CBC after the window filtering algorithm is opened in the closed loop. The inset shows that the phase difference of the two pulsed beams in the phase-locking process.

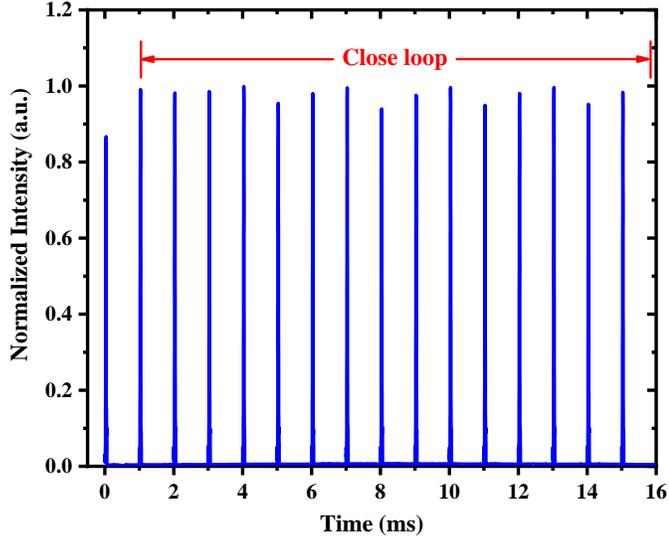

Fig. 7. Coherent combination effect of the two pulsed beams in the time-domain during the phase-locking process.

It can be seen from Fig. 6 that after the pulses are filtered by self-designed window filter algorithm, the multi-dithering phase lock algorithm can be successfully used to compensate the phase difference of the two pulsed beams to 0 within 1 ms (Illustration in Fig. 6), and the average light intensity after the combination of two pulsed beams becomes stable from the original large fluctuation, and the final light intensity increases to the ideal value (0.9 $I_{max}$). From the analysis of the phase-locking process, when the window filter algorithm starts, the phase difference of the two beams is relatively large, as shown in the inset in Fig. 6. Correspondingly, it can be seen from Fig. 7 that the first pulse is not at maximum value. With the operation of the algorithm, the phase difference between the two pulsed beams is rapidly reduced to near 0, and the pulse sequence also tends to near the maximum value. Therefore, it can be seen that the adaptive window filter algorithm we designed can filter out the pulses whose repetition frequency and phase noise frequency are in the same order of magnitude, and the detection and filtering of the pulses takes a short time (0.1 ms). After filtering the pulses in the phase noise, a fast (1 ms) and stable phase lock can be achieved by multi-dithering algorithm.

## IV. CONCLUSION

This paper proposes a new method of detecting, identifying, and filtering pulse noise. According to the autocorrelation characteristics of pulse signals, a set of window filtering algorithms is designed by itself, which can effectively filter the pulse signal doped in phase noise within 0.1ms. Compared with the traditional band-pass filter or low-pass filter with appropriate cut-off frequency to filter out pulses, the complexity of the phase-locked system is reduced, and a lot of time and energy spent to adjusting the experimental system are reduced. Secondly, the window filter algorithm we designed only interpolates the pulse noise pollution points, does not change the phase noise value of the non-polluted points, which can well protect the original phase noise. The simulation results show that after the pulses are filtered out when the multi-dithering algorithm is applied, the phase difference of the two pulsed signal lights can be compensated to 0 within 1ms, and the closed-loop phase-locked coherent combination of the two pulsed beams can be realized, and the phase is corrected. The phase correction times are few, the phase lock effect is more stable, and the final light intensity increases to the ideal value (0.9 Imax). In addition, the adaptive window filtering algorithm we proposed can also be extended to more channels of LRF pulse optical coherent combination. In the future, we plan to use experimental methods to further research and verify the algorithm, which is expected to be applied to an LRF large array fiber laser coherent combination system, and further lay the foundation for fiber phased array lidar.